\documentclass[twocolumn,floatfix,showkeys,showpacs,prd,aps,tightenlines,letterpaper,superscriptaddress, nofootinbib]{revtex4-1}
\usepackage{lineno}
\usepackage[utf8]{inputenc}

\usepackage{amsmath}
\usepackage{graphicx}
\usepackage{float}
\usepackage{subfigure}
\usepackage{psfrag}
\usepackage{verbatim} 
\usepackage[usenames]{color}

\usepackage{dcolumn}
\usepackage{bm}
\usepackage{longtable}
\usepackage[mathscr]{eucal}
\usepackage{mathrsfs}
\usepackage{xspace}
\usepackage{tikz}
\usetikzlibrary{arrows,shapes,trees,decorations.pathreplacing}
\usepackage{import}
\usepackage{svn-multi}
\usepackage{leftidx}
\usepackage{hyperref}
\usepackage{gensymb}
\hypersetup{
  colorlinks=true,
  citecolor=blue,
  linkcolor=blue,
  filecolor=blue,   
  urlcolor=blue,
}

\newcommand{\msun}{\ensuremath{\mathrm{M}_{\odot}}}

\usepackage{enumitem}
\usepackage{etoolbox}

\newcommand{\orcidicon}[1]{\href{https://orcid.org/#1}{\includegraphics[height=\fontcharht\font`\B]{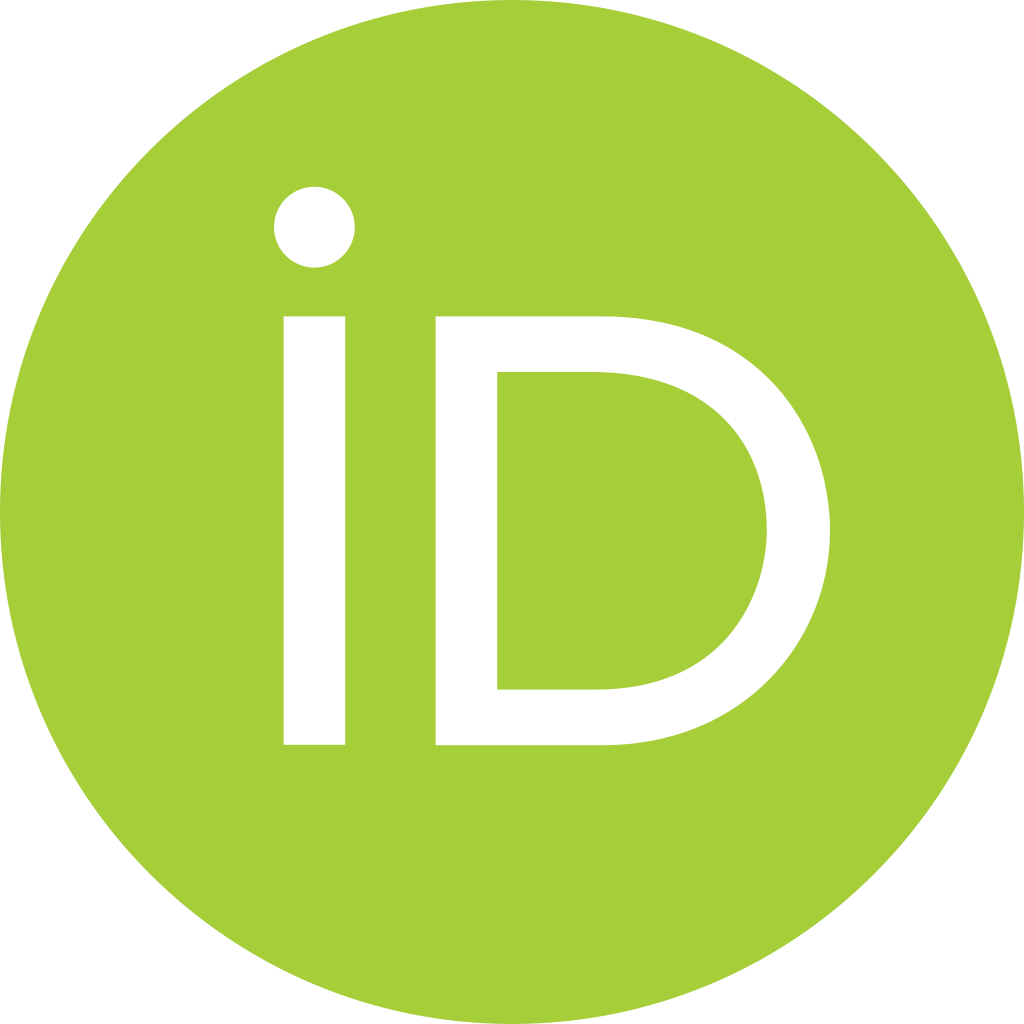}}}

\makeatletter

\patchcmd\frontmatter@PACS@format{\addvspace{11\p@}}{}{}{}
\pretocmd\frontmatter@keys@format{\addvspace{11\p@}}{}{}
\patchcmd{\titleblock@produce}
  {\@pacs@produce\@pacs\@keywords@produce\@keywords}
  {\@keywords@produce\@keywords\@pacs@produce\@pacs}
  {}{}
\makeatother

\begin{document}
\title{Flexible and Fast Estimation of Binary Merger Population Distributions with Adaptive KDE}

\author{Jam~Sadiq \orcidicon{0000-0001-5931-3624}}
\email{contact: jam.sadiq@usc.es}
\affiliation{Instituto Galego de F\'{i}sica de Altas Enerx\'{i}as, Universidade de Santiago de Compostela, Santiago de Compostela, Galicia, Spain}

\author{Thomas~Dent \orcidicon{0000-0003-1354-7809}}
\affiliation{Instituto Galego de F\'{i}sica de Altas Enerx\'{i}as, Universidade de Santiago de Compostela, Santiago de Compostela, Galicia, Spain}

\author{Daniel~Wysocki \orcidicon{0000-0001-9138-4078}}
\affiliation{University of Wisconsin--Milwaukee, Milwaukee, Wisconsin, USA}

\begin{abstract} 
The LIGO Scientific, Virgo and KAGRA Collaborations recently released the third gravitational wave transient catalog or GWTC-3, significantly expanding the number of gravitational wave (GW) signals. To address the -- still uncertain -- formation channels of the source compact binaries, their population properties must be characterized.  The computational cost of the Bayesian hierarchical methods employed thus far scales with the size of the event catalogs, and such methods have until recently assumed fixed functional forms for the source distribution. Here we propose a fast and flexible method to reconstruct the population of LIGO--Virgo merging black hole (BH) binaries without such assumptions. For sufficiently high event statistics and sufficiently low individual event measurement error (relative to the scale of population features) a kernel density estimator (KDE) reconstruction of the event distribution will be accurate. We improve the accuracy and flexibility of KDE for finite event statistics using an adaptive bandwidth KDE (awKDE). We apply awKDE to publicly released parameter estimates for 44 significant (69) BH binary mergers in GWTC-2 (GWTC-3), in combination with a fast polynomial fit of search sensitivity,
to obtain a non-parametric estimate of the mass distribution, and compare to Bayesian hierarchical methods. We also demonstrate a robust peak detection algorithm based on awKDE and use it to calculate the significance of the apparent peak in the BH mass distribution around $35\,\msun$. We find such a peak is very unlikely to have occurred if the true distribution is a featureless power-law (significance of $3.6\sigma$ for confident GWTC-2 BBH events, $3.0\sigma$ for confident GWTC-3 BBH events).  
\end{abstract}


\maketitle

\section{Introduction}

Advanced LIGO's and Advanced Virgo's \cite{LIGOScientific:2014pky,VIRGO:2014yos} first three observing runs have produced many tens of confident detections of binary compact object mergers via gravitational wave (GW) emission, catalogued in GWTC-1 \cite{LIGOScientific:2018mvr}, GWTC-2 \cite{LIGOScientific:2020ibl}, GWTC-2.1 \cite{LIGOScientific:2021usb}, and GWTC-3 \cite{LIGOScientific:2021djp}.\footnote{Each successive catalog update includes previous detections.} A detailed investigation of the population properties of binary black hole (BBH) mergers, the most commonly detected source type, has been published in \cite{Abbott:2020gyp} focusing on several population characteristics including their component masses and spins using GWTC-2 \cite{LIGOScientific:2020ibl} events. This investigation was updated recently in \cite{LIGOScientific:2021psn} for GW observations up to the end of the O3 run, as catalogued in GWTC-3 \cite{LIGOScientific:2021djp}, where updated events reported in GWTC-2.1 \cite{LIGOScientific:2021usb}, are included.

One objective has been to reconstruct the primary mass distribution of merging BBH, in order to address questions in stellar evolution, BH formation and binary formation channels. The primary BH masses have lower measurement uncertainty compared to other parameters such as binary spins, thus their distribution is expected to yield significant information for eventual comparison with astrophysical model predictions. The binary chirp mass $\mathcal{M} \equiv (m_1m_2)^{3/5}(m_1+m_2)^{-1/5}$, where $m_{1,2}$ are the source binary masses, is also measurable with (relatively) high precision, thus its distribution is comparable with model predictions stated in terms of binary properties  \cite[e.g.][]{Broekgaarden:2021efa,Dominik:2014yma}.

\begin{figure}
\vspace*{-0.2cm}
\centering
\includegraphics[width=0.98\linewidth]{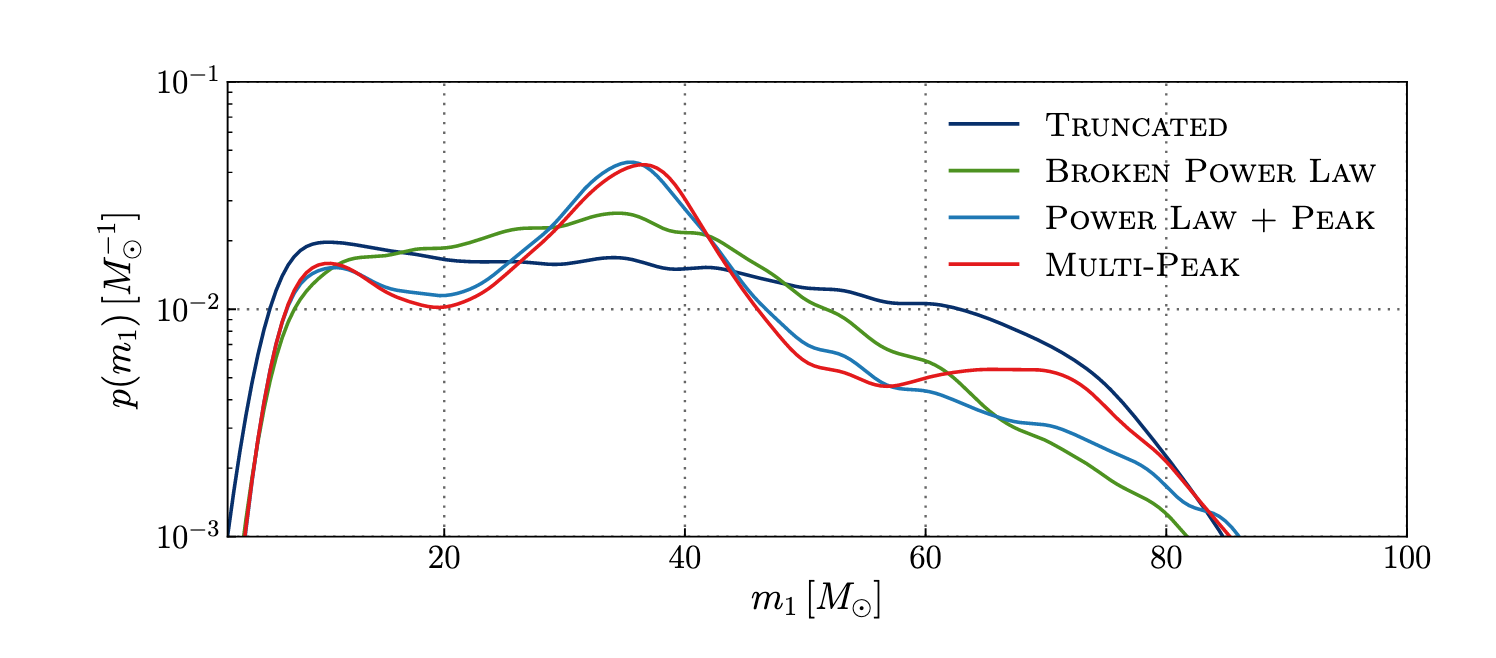}
\caption{Observed primary black hole mass distribution predicted by each mass model in \cite{Abbott:2020gyp}; reproduced from that publication.}
\label{fig:pop_paper_models}
\end{figure}

In \cite{Abbott:2020gyp} specific functional forms and Bayesian hierarchical analysis are used to infer the mass distribution of binary mergers from these observed gravitational wave events: see Figure~\ref{fig:pop_paper_models} for one output of such inference. These methods are well established, but as our observed sample grows with future observing runs, both computational cost and modeling complexity will increase accordingly. An alternative approach to reconstructing the population is non-parametric methods that allow deviations from a predetermined functional form to better fit observations, or that can approximate arbitrary functional forms.  
Non-parametric methods implemented via Bayesian analysis have recently been described in \cite{Tiwari:2020otp,Tiwari:2020vym, Veske:2021qis, Rinaldi:2021bhm, Edelman:2021zkw,  Powell:2019nmw, Farr:2017gtv}; the motivation of this work is similar, but we prioritize computational simplicity and speed.

In this paper we propose a fast and flexible method for reconstructing binary merger population distributions via \textit{adaptive bandwidth kernel density estimation} (KDE), using a publicly available code \cite{awkdecodelink}. This method addresses the same goals as the hierarchical analyses mentioned above, but without assuming a particular functional form of the distribution beyond the choice of a kernel function (here, the choice of a Gaussian kernel enforces a smooth estimate). A standard Gaussian KDE with a fixed (global) bandwidth is unlikely to give a accurate representation of the BBH mass distribution, due to its complexity; \cite{Mandel:2009pc} already identified the risk of significantly biased population estimates due to inappropriate or inflexible bandwidth choice. 

Here, we deploy an adaptive KDE (awKDE) estimation method in combination with cross-validation to address the bandwidth problem: specifically, to ensure a locally appropriate bandwidth across a broad parameter space with a wide dynamic range of densities.
We estimate the mass distribution of binary mergers from observed events, enabling us to validate and check the model assumptions of the standard Bayesian analyses in a flexible and computationally efficient way; this nonparametric approach can also potentially discover features not described by parametric models. The awKDE estimate of the detected event distribution also yields a differential merger rate estimate, using a fit of the search sensitivity as in \cite{2019dan}. We also compute an uncertainty estimate (confidence interval) for the population KDE using the \textit{bootstrap} technique \cite{bootstrapcitation}. 

There is an apparent peak around $35M_{\odot}$ in the primary mass distribution of observed GW events, as seen in the \textsc{Power Law + Peak} \cite{Talbot:2018cva} and \textsc{Multi Peak} model results in \cite{Abbott:2020gyp} (see Fig.~\ref{fig:pop_paper_models}).  In this work we also propose a method to determine the significance of one or more prominent peaks in the detected distribution, quantifying the probability that comparable statistical fluctuations might occur in the presence of selection effects. We thus introduce a new peak detection algorithm \cite{peakdetectioncodelink} and apply it to GWTC-2 events.  Investigating the presence and significance of such peaks 
will further help towards understanding BH binary formation; for instance, it has been suggested that pair-instability dynamics in supernovae could induce similar features in the BH mass function \cite[recent discussions include][]{Farmer:2019jed, Baxter:2021swn, Woosley:2021xba}; however, as yet no definite signature has been identified.

The paper is organized as follows: In Section II we describe the motivation and reconstruction of the population probability density function using the awKDE algorithm and its application to GW detections, including astrophysical merger rates.  We further describe our procedure for computing the uncertainty estimates of our distribution function using bootstrap and awKDE Gaussian kernel contributions.  In Section III we apply awKDE to reconstruct the BBH mass distribution from the observed LVK events. In Section IV we describe a peak detection algorithm to determine the significance of features beyond a power law distribution, and obtain results for GWTC-2 and GWTC-3 detected events.  In Section V we summarize our conclusions.

\section{Reconstruction Method}

We propose the use of a kernel density estimate with adaptive bandwidth selection for reconstructing the probability distribution of source parameters for compact binary mergers observed via GW. This method is non-parametric, straightforward to apply, and enables the identification of general features in the distributions that may be an important input in the astrophysical interpretation of the merging binary population. 

\begin{figure}
\centering
\includegraphics[width=0.92\linewidth]{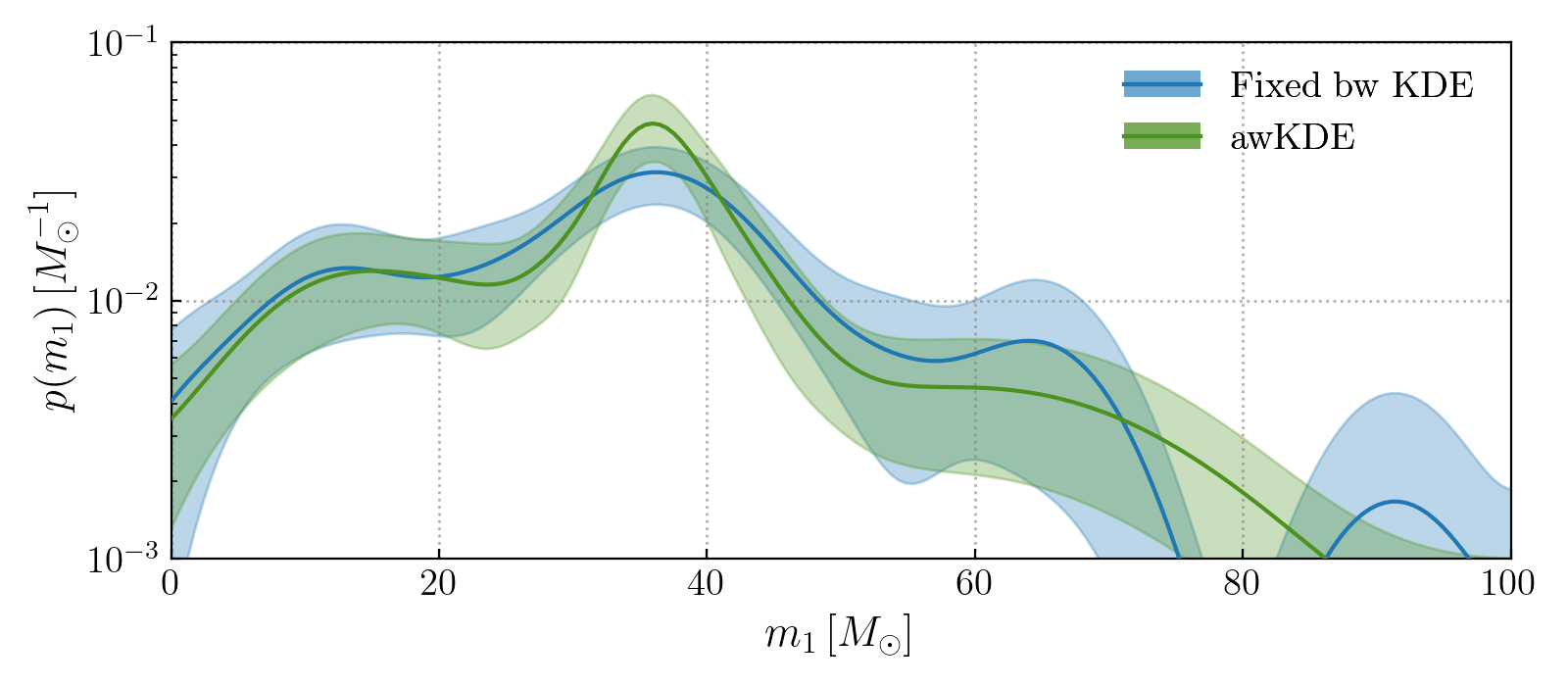}
\caption{Fixed global bandwidth versus awKDE reconstruction of primary BH masses using GWTC-2 detected events.  
}
\label{fig:compare_fixedbw_vs_awKDE}
\end{figure}

While a KDE with fixed (global) bandwidth, chosen for example via Silverman's rule \cite{Silverman1986}, is known to be a suitable method to estimate a distribution close to a single Gaussian, the mass (and possibly also spin) distributions of BBH mergers appear to have a more complex structure, which we do not expect to be well reconstructed by a simple KDE. 
In particular, the primary and secondary mass distributions may be composed of several components with widely differing mass scales and densities (which have in some cases, e.g.~\cite{Talbot:2018cva}, been modelled by power laws and Gaussian peaks).  Thus, we consider an extension where the bandwidth varies locally according to an initial estimate of the density of sample points \cite{TerrellScott, SainScott}. The differences between a fixed global bandwidth KDE and our proposed awKDE are illustrated in Figure~\ref{fig:compare_fixedbw_vs_awKDE}. 

The fixed bandwidth KDE overestimates the width (underestimates the height) of the observed peak around $35\,\msun$, but also yields a probably unphysical gap in the estimated distribution around $80\,\msun$; these undesirable aspects arise because a fixed bandwidth cannot both reconstruct small-scale features in regions with a high density of points, and supply enough smoothing to avoid artefacts in low density regions. 
 
An adaptive bandwidth KDE \cite{wang2011bandwidth} is implemented in the open source code \textsc{awkde} \cite{awkdecodelink}.  We first describe the construction of a KDE from observations $X_i$, $i = 1 \ldots n$: for instance, $X_i$ may be a measured property of a binary merger. Later, we will describe how measurement uncertainties in these individual event properties are incorporated. 

The algorithm computes a density estimate $\hat{f}$ via
\begin{equation}\label{awkde_initial}
    \hat{f}(x) = n^{-1} \sum_{i=1}^{n} \frac{1}{h \lambda_i} K \left( \frac{x- X_i}{h \lambda_i} \right),
\end{equation}
where $K(\cdot)$ is the standard Gaussian kernel,
\begin{equation}
    K(z) = \frac{1}{\sqrt{2\pi}} \exp\left(\frac{-z^2}{2}\right),
\end{equation}
$n$ is the total number of samples and, in general, the product $h \lambda_i$ takes the role of a local bandwidth with $h$ being the global bandwidth.

The first step 
is the computation of a pilot estimate $\hat{f}_0$ setting $\lambda_i = 1$ for all $i$, which is a standard fixed bandwidth KDE.  Based on the pilot density $\hat{f}_0$, the local bandwidth accounting for variations in the density of samples is obtained via 
\begin{equation}
    \lambda_i  = \left( \frac{\hat{f}_0(X_i)}{g} \right)^{-\alpha},
\end{equation}
where $\alpha$ is the local bandwidth sensitivity parameter ($0 < \alpha \leq 1$) and $g$ is a normalization factor
\begin{equation}
    \log g = n^{-1} \sum_{i=1}^{n} \log \hat{f}_0(X_i).
\end{equation}
Finally the adaptive KDE $\hat{f}(x)$ is obtained by evaluating \eqref{awkde_initial} with the variable (local) bandwidth $h\lambda_i$.  
The above method assumes one-dimensional data $X_i$; the method may also be applied to two- or more-dimensional data by linearly transforming the data to have zero mean and unit covariance.

The method requires a choice of the initial global bandwidth $h$ and sensitivity parameter $\alpha$: we use the \textit{leave-one-out cross-validation} method \cite{hastie01statisticallearning} to determine these values.
As a figure of merit for the cross-validation we use the (log) likelihood,
\begin{equation}
  \log \mathcal{L}_{\rm LOO} = \sum_{i=1}^{n} \log \hat{f}_{{\rm LOO},i}(X_i),
\end{equation}
where $\hat{f}_{{\rm LOO},i}$ is the KDE constructed from all samples \emph{except} $X_i$.  This choice, as it is linear in the logarithm of the estimate at observed values, will penalize \emph{relative} errors. Since we wish to obtain an accurate estimate of densities over a large dynamic range, the log likelihood is more suitable than FOM based on absolute error or squared absolute error. We employ a grid search over a range of $h$ and $\alpha$ values; however, we often find the likelihood is maximized with $\alpha$ at or close to $1$, thus in some cases (discussed later) we will impose $\alpha = 1$ rather than conduct the full 2d grid search.

\subsection{Application to GW observations}
\label{sec:gw_recon}

The component masses of observed GW binary mergers have significant measurement uncertainties, which we wish to incorporate in reconstructions of the mass distribution.  
The parameters of each binary are obtained by Bayesian inference using models of the emitted GW waveform \cite[e.g.][]{Veitch:2014wba}, referred to as parameter estimation (PE).  We consider detected mergers labelled by $i = 1 \ldots n$: uncertainties in a given parameter $X$ are quantified via random values $X_i ^{k}$ drawn from the posterior of the $i$th merger.  Typically, some thousands of parameter samples are available per merger event \cite{GWTC2PEdata}.\footnote{Note that we use samples obtained with a distance prior proportional to the comoving volume element times (redshifted) source time.}

Although we use a random selection of PE samples for the KDE including measurement uncertainty, to first obtain the optimum global bandwidth and sensitivity parameter $\alpha$ we evaluate the likelihood $\mathcal{L}_{\rm LOO}$ using only the \emph{median} parameter value for each merger as the data $X_i$.  This choice reduces the computational cost and avoids over-fitting of random fluctuations: note that these medians are independent values, whereas the PE samples for a given event $i$, $X_i^{k}$, are not independent of one other, as they are all correlated with the (unknown) true source mass: thus, the PE samples for a given event are not independent draws from the population distribution that the KDE aims to reconstruct.

Having obtained the optimal $h$ and $\alpha$ choices, we construct the population KDE using 100 randomly chosen samples for each event. We verify that increasing the number of PE samples did not change our results significantly: considering KDE outputs using 10, 30, 100, and 300 PE samples and taking the 300-sample fit as a baseline, we find the maximum fractional difference, $\max_{m_1} f_{300}(m_1)^{-1}[f_{n}(m_1) - f_{300}(m_1)]$, for the 10-, \mbox{30-,} and 100-sample outputs to be $\sim20\%$, $\sim15\%$, and $\sim6\%$, respectively.  We note though that for large measurement errors relative to the scale of structures in the underlying distribution, the KDE result is likely to be over-dispersed, i.e., any sudden variations in the actual density will be smoothed out.\footnote{In principle this bias can be tackled by the computationally demanding hierarchical methods, if they model functional forms that are sufficiently close to the true distribution.}

Our major source of uncertainty for the population KDE lies in the finite number of binary merger events, and the resulting count fluctuations in the density estimate at a given parameter value.  We estimate this uncertainty  by bootstrap resampling \cite{bootstrapcitation} over the merger events $i$: i.e., we create a large number of bootstrapped data sets, each containing some number of copies of the PE samples for each event, with the number of copies following a binomial distribution.  For the cases we consider, this distribution is well approximated by the Poisson distribution with unit mean, hence we expect this procedure to give the correct scaling of uncertainties due to event count fluctuations. We extract the median and 5th and 95th percentiles from 1000 bootstrap iterations to obtain the confidence interval at any given parameter value; by plotting all bootstrap KDEs we can also visually identify regions of high and low uncertainty.

\subsection{Fast approximate uncertainty estimate}
\begin{figure}
    \centering
    \includegraphics[width=.98\linewidth]{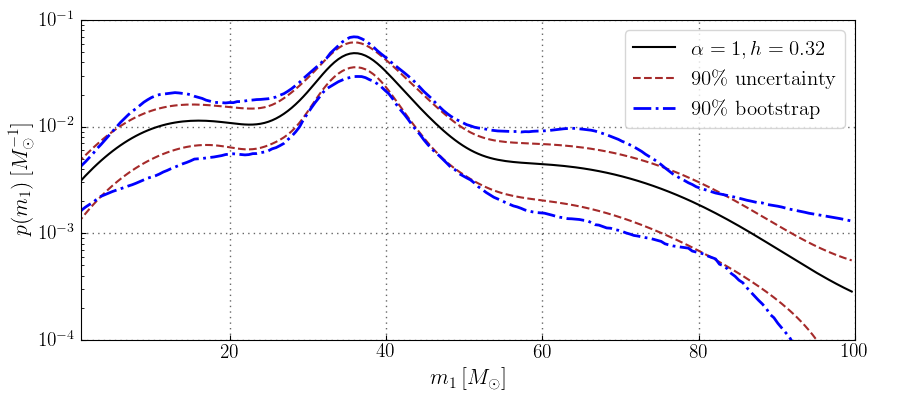}
    \caption{awKDE (black solid line) for GWTC-2 events with error estimates using a standard bootstrap method (blue dot-dashed lines) and from the Gaussian Kernel coefficients from awKDE code (brown dashed lines)  using Eq.~\eqref{eq:error_awkde}.  The two error estimates are in good agreement for most parameter values.}
    \label{fig:compare_error_estimates}
\end{figure}

While constructing an awKDE with median parameter values from PE, or in general using one parameter value per event, we can obtain a rapid uncertainty estimate by considering the awKDE Gaussian kernel contributions due to observed events at any given parameter value $x$ (accessible with our modified awKDE algorithm \cite{awkdecodelink}).  These contributions or KDE coefficients are
\begin{equation} \label{eq:kde_coeff}
    c_k(x) = n^{-1} \frac{1}{h \lambda_k}
    K\left( \frac{x - X_k}{h \lambda_k} \right),
\end{equation}
where $k$ labels the observed events. 
We may estimate the standard deviation of the coefficients, $\sigma_{c}$, as 
\begin{equation} \label{eq:kde_sigma}
    \sigma_{c}^2(x) \simeq \langle c_k ^2\rangle - \langle c_k\rangle^2,
\end{equation}
where $\langle \cdot \rangle$ represents the mean over observed events.  As the KDE is the sum of $n$ such coefficients its variance is a factor $n$ larger, thus we take $\sigma_{\rm KDE}(x) = \sqrt{n}\sigma_{c}(x)$.  
The resulting estimated 90\% confidence interval is $1.64\,\sigma_{\rm KDE}(x)$ above and below the central value,\footnote{The factor $1.64$ corresponds to the 95th percentile of the standard normal distribution.} i.e.\
\begin{equation}
    \epsilon_\mathrm{KDE}(x) = 1.64 \sqrt{n}\sigma_c(x). 
\label{eq:error_awkde}
\end{equation}
We compare this estimate with the bootstrap uncertainty region and find them in a good agreement for most parameter values, as shown in Figure~\ref{fig:compare_error_estimates}.  Clearly this direct variance estimate is computationally much more efficient than the bootstrap.

While this coefficient-based estimate agrees with the bootstrap for small bandwidths (relative to the range of sample values $X_i$), for large values of bandwidth it becomes significantly smaller than the bootstrap uncertainty. We also see from Fig.~\ref{fig:compare_error_estimates} that the coefficient-based estimate tends to underestimate the bootstrap uncertainty in regions of low density. A complete explanation of this behaviour should include more complicated effects due to the two-step nature of the adaptive KDE: for instance, the estimate of \eqref{eq:error_awkde} does not account for any systematic variation in coefficients due to changes in the bandwidths $\lambda_k$ between different data realizations.  Also, the coefficient-based estimate considers only the variance of the density, which does not account for uncertainty distributions that deviate significantly from Gaussian, such as Poisson statistics in the small number regime.

We may get some insight from heuristic arguments: if we approximate the Gaussian kernel coefficients at a given $x$ as constants $c_L$ (`large' coefficients) for $n_L$ data points $X_k$ `close' to $x$ relative to their bandwidths $h\lambda_k$, and vanishing for the remaining $n-n_L$ points `far away' from $x$, then 
$\sigma_{\rm KDE}(x)$ becomes approximately $(c_Ln_L/n)\sqrt{n_L^{-1} - n^{-1}}$.  This corresponds to the variance of a binomial distribution with a success rate $n_L/n$.  When the bandwidths are large, $n_L$ will be close to $n$ (many events will have large and near-equal coefficients), leading to a cancellation.  

However, given that the data are produced by an underlying Poisson process, as the total number of astrophysical events observed is not fixed, we also considered a modified estimate which scales with the variance of the count of `close' events, reducing to $(c_Ln_L/n)\sqrt{n_L^{-1}}$.  This corresponds to removing the second term in Eq.~\eqref{eq:kde_sigma}, i.e.\ 
%
\begin{equation}\label{eq:sigmahat}
  \hat{\sigma}^2_c(x) = \langle c_k^2 \rangle.
\end{equation}
We find that the $90\%$ error $\hat{\epsilon}_\mathrm{KDE}(x)$  corresponding to this modified estimate agrees much more closely with bootstrap uncertainties for large bandwidths.  This property motivates our use of $\hat{\epsilon}_\mathrm{KDE}$ in the peak detection method to be introduced in Section~\ref{sec:peak}.

\section{awKDE population reconstruction from LIGO-Virgo detections}

In this section we apply the methods described above to reconstruct the distribution of parameters for BBH observed so far in the advanced detector runs.  We divide this section into results obtained using the GWTC-2 event set, comprising events from the O1, O2 and O3a runs 
\cite{LIGOScientific:2018mvr,  LIGOScientific:2020ibl}, which we compare to the collaboration results in \citet{Abbott:2020gyp}; and results using the GWTC-3 set \cite{LIGOScientific:2021djp} (including an updated release for data and events up to O3a, called GWTC-2.1 \cite{LIGOScientific:2021usb}, where available). 

\subsection{BBH population from GWTC-2}

\begin{figure}[t]
\centering
\includegraphics[width=.98\linewidth]{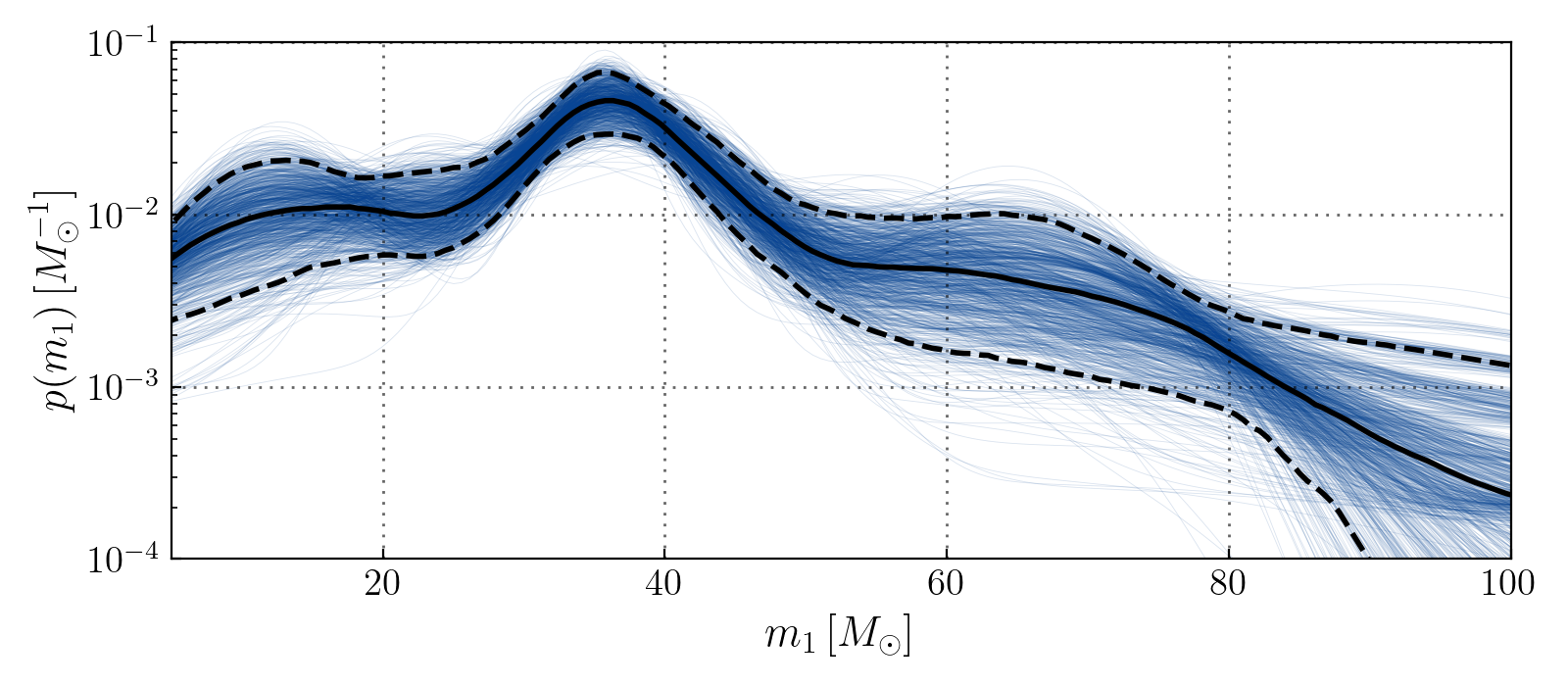}
\includegraphics[width=.98\linewidth]{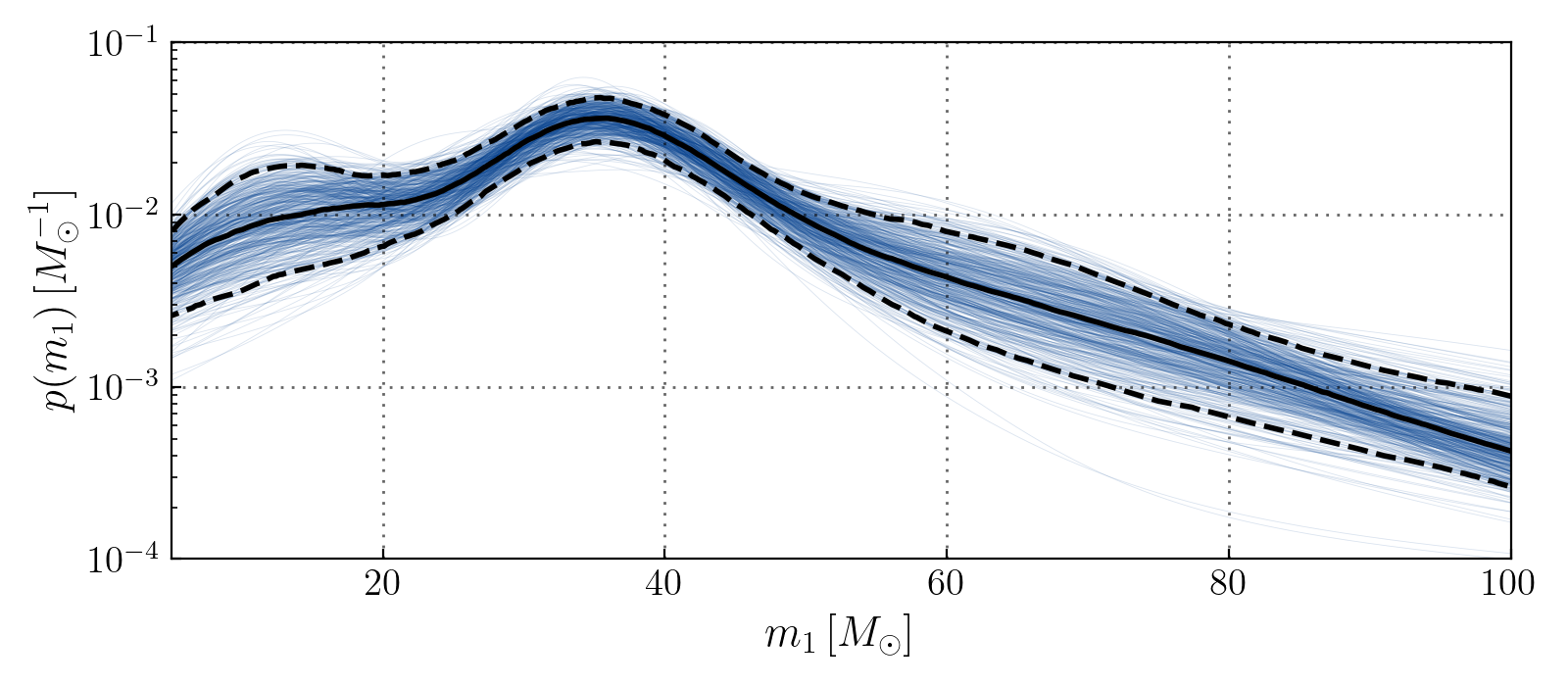}
\caption{KDE with uncertainty estimates of source frame primary mass, $m_1$, for detected BBH events in the O1, O2 and O3a observing runs. 
Top: KDE using the PE sample median for each event. Bottom: KDE constructed using 100 random samples from each observed event. The median (black solid), 90\% confidence interval (black dashed lines) and blue curves are constructed using a bootstrap. Our KDE results match closely the \textsc{Power Law + Peak} \cite{Talbot:2018cva} and \textsc{Multi Peak} model results in \cite{Abbott:2020gyp}, see Fig.~\ref{fig:pop_paper_models}.
\label{fig:m1kde}
}
\end{figure}

We use detected GW events from GWTC-2 
\cite{LIGOScientific:2018mvr,  LIGOScientific:2020ibl} to reconstruct the KDE of parameters for BBH observations.  As in \cite{Abbott:2020gyp}, we consider significant events (false alarm rate below $1/$yr) for which both component masses are above $3\,\msun$, finding 44 such mergers. 
We first consider the primary mass $m_1$ defined in the binary source frame. 
The resulting KDE bootstrap samples, median estimate and 90\% confidence intervals are shown in Figure~\ref{fig:m1kde}, where the top panel uses only the median $m_1$ value for each binary merger and the lower panel accounts for measurement uncertainties via the PE samples. 
As described in Sec.~\ref{sec:gw_recon}, the KDE bandwidth and sensitivity parameter $\alpha$ are optimized via leave-one-out cross validation using median $m_1$ sample values, and the same optimized parameters are used to evaluate the KDE including PE sampling of mass uncertainties.

The resulting primary mass distributions match well those inferred using parametric models in \cite{Abbott:2020gyp}, specifically with the \textsc{Power Law + Peak} \cite{Talbot:2018cva} and \textsc{Multi Peak} models most preferred by the GWTC-2 data.  We note that there is a clear global maximum in the observed distribution at $\sim 35\,\msun$ and that the uncertainty in the mass distribution is smallest here.  The awKDE is also able to reconstruct a wide dynamic range of densities without excessive statistical uncertainties. 

\begin{figure}[!h]
\centering
\includegraphics[width=.98\linewidth]{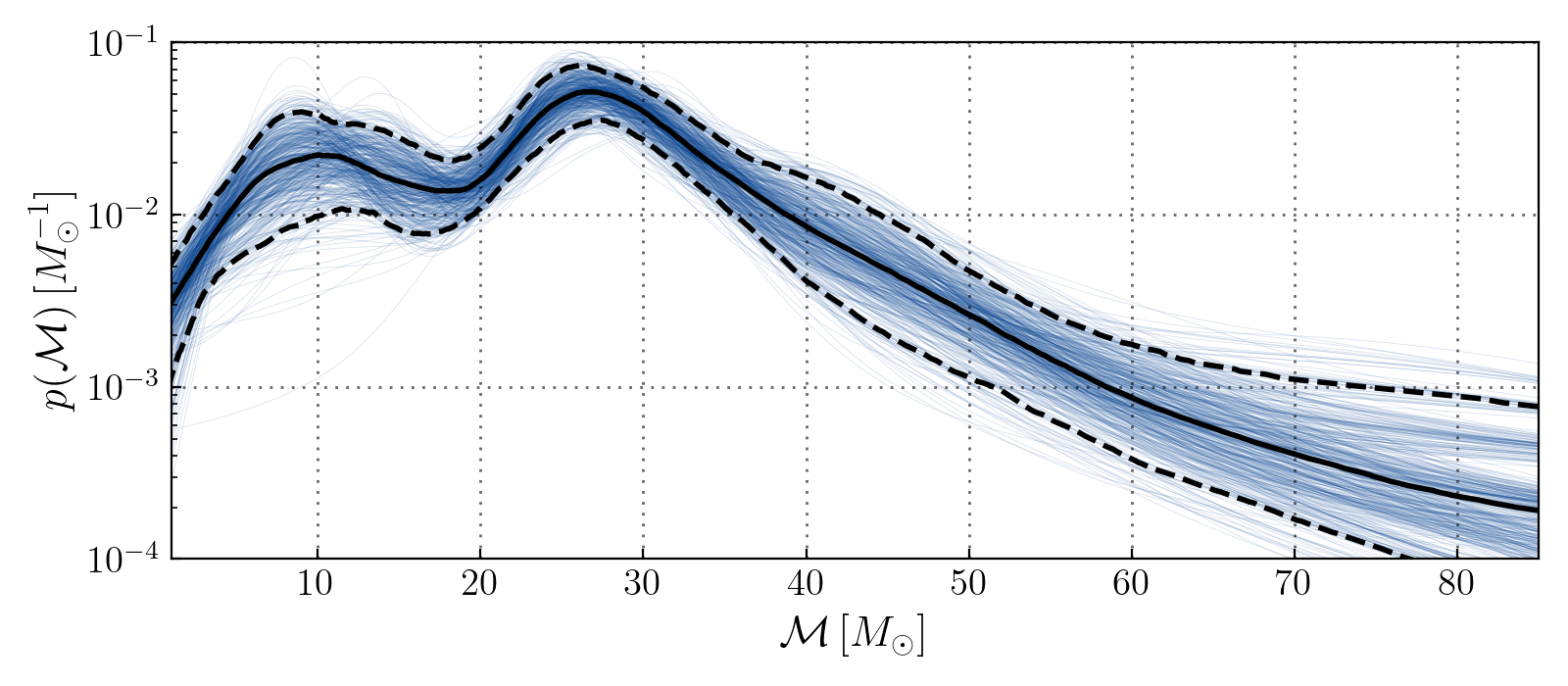}
\caption{KDE with uncertainty estimates of source chirp mass, $\mathcal{M}$,
for detected events in the GWTC-2 or O1, O2 and O3a observing runs. KDE constructed using 100 random samples from each observed event. The median (black solid), 90\% confidence interval (black dashed lines) and blue curves are constructed using a bootstrap.  In addition to the principal peak in the distribution around 25--30\,\msun, there is a hint of further structure around 10\,\msun.}
\label{fig:Mchirpkde}
\end{figure}

As an alternative to the primary mass, we also consider the chirp mass $\mathcal{M}$ for observed BBH mergers, which is also measured with relatively small uncertainty. 
The corresponding KDEs are shown in Figure~\ref{fig:Mchirpkde}.
Here in addition to the expected peak just below $30\,\msun$ we note hints of a secondary peak around $\mathcal{M}\simeq 10$, though small compared to the estimated uncertainties: however, with this data we are not able to distinguish the multiple peaks claimed in \cite{Tiwari:2020otp} from random fluctuations.

\paragraph{Merger rate estimation from awKDE}
\label{sec:rateest}
We also consider estimation of BBH merger rates using our awKDE results. The additional ingredient in this analysis is the sensitive volume-time $VT$ over which a source is detectable: we quantify this based on an approximation for the sensitivity of GW detectors, with corrections to account for the actual behaviour of searches in real data~\cite{2019dan}.  
The idea is to calibrate a semi-analytic function $VT_{\rm analytic}$ against the re-weighted result of injections (i.e.\ simulated signals added to real data and analyzed by search pipelines) $VT_{\rm inj}$, assuming a parameterized relationship between them that can be expressed via basis functions.  

\begin{figure}[th]
\centering
\includegraphics[width=.99\linewidth]{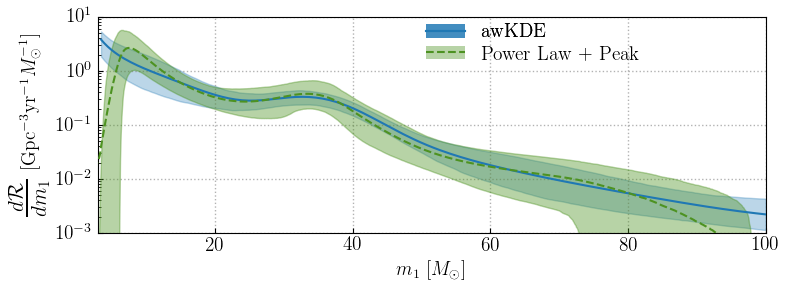}
\caption{Rate estimates using adaptive KDE and sensitive volume for BBH events in GWTC-2. The green band curve is the rate estimate using the \textsc{Power Law + Peak} model  
from \cite{Abbott:2020gyp}; the blue band shows a rate estimate using KDE results from median PE values and a fit to search sensitivity.
}
\label{fig:rateestimate}
\end{figure} 

For the GWTC-2 observing runs, the `uncalibrated' semi-analytic sensitivity is estimated using the criterion that the signal-to-noise ratio (SNR) of a signal in the second most sensitive detector, using specific reference power spectral densities, should be greater than 8.  This semi-analytic estimate $VT_{\rm analytic}$ is obtained for a given observing time $T$ over a grid of intrinsic source parameters such as component masses. For the injection $VT_{\rm inj}$, one performs a set of injections and counts the number detected by search pipelines to obtain an average $VT_{\rm inj}$ for the injected population at given intrinsic parameters. The calibrated estimate of $VT$ is then obtained by fitting a correction function to the deviations of $VT_{\rm inj}$ from $VT_{\rm analytic}$, as described in \cite{sensitivityDan} and applied in \cite{Abbott:2020gyp}.

We compute the merger rate density using this corrected $VT$ and our KDE results for the primary mass $\hat{f}(m_1)$ from observations up to O3a.  We obtain $VT(m_1)$ assuming a power law distribution for the secondary mass $m_2$, taking the median power $\beta_q = 1.26$ from the \textsc{Truncated} model samples from \cite{Abbott:2020gyp}, and find the merger rate density via 
\begin{equation}
\frac{dR}{dm_1} = n \frac{\hat{f}(m_1)}{VT(m_1)}.
\end{equation}
Our estimate is in good agreement with the results in \cite{Abbott:2020gyp} for the \textsc{Power Law + Peak} model, as shown in Fig.~\ref{fig:rateestimate}.

\subsection{BBH population from GWTC-3}

We also used recently available public data from GWTC-3 \cite{LIGOScientific:2021djp} to construct the distribution of primary and chirp masses using PE samples for 69 confident BBH events with a false alarm rate
below $0.25/$yr.  Note that this data set includes a small number of events identified in a reanalysis of O3a data with improved search pipelines, detailed in GWTC-2.1~\cite{LIGOScientific:2021usb}. 
We find new features in the distributions, consistent with the results in \cite{LIGOScientific:2021psn}. 
\begin{figure}[!th]
    \centering
    \includegraphics[width=.95\linewidth]{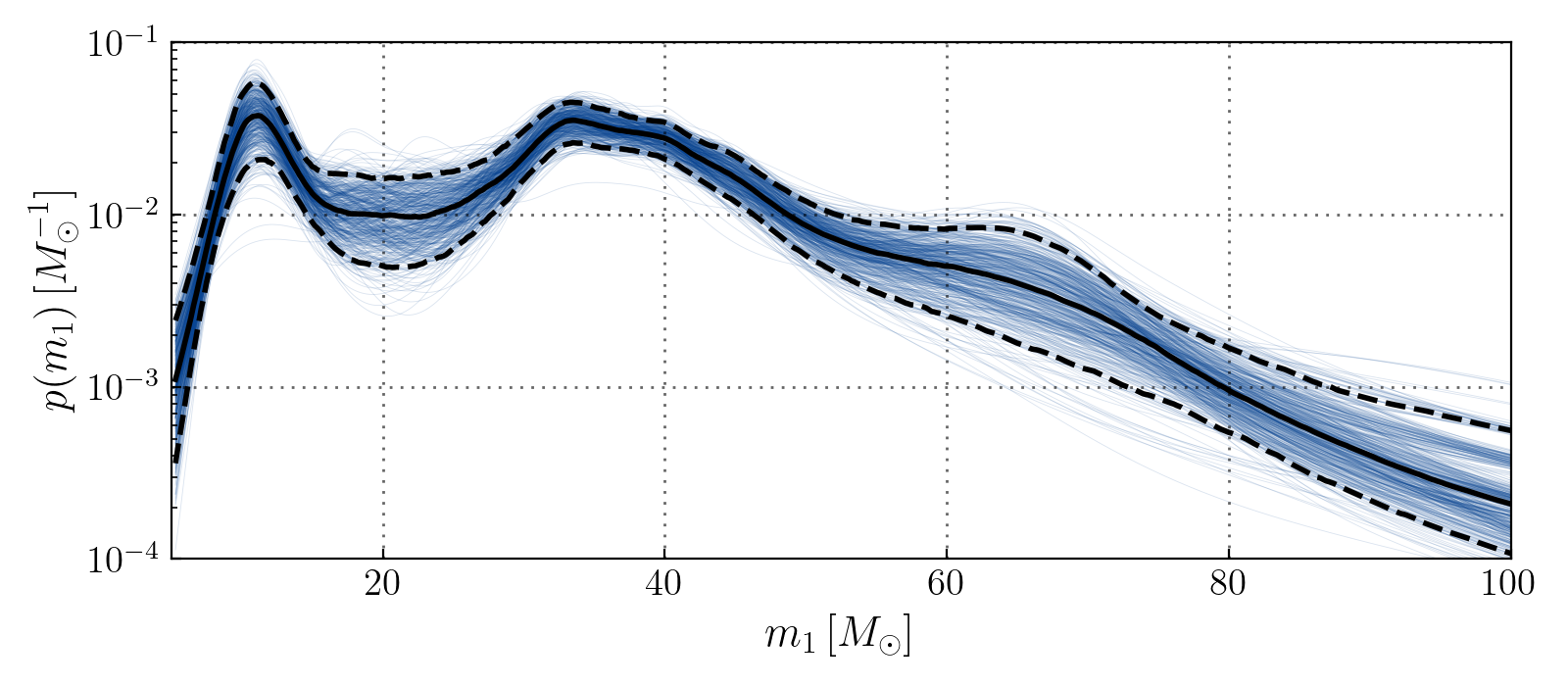}
    \includegraphics[width=.95\linewidth]{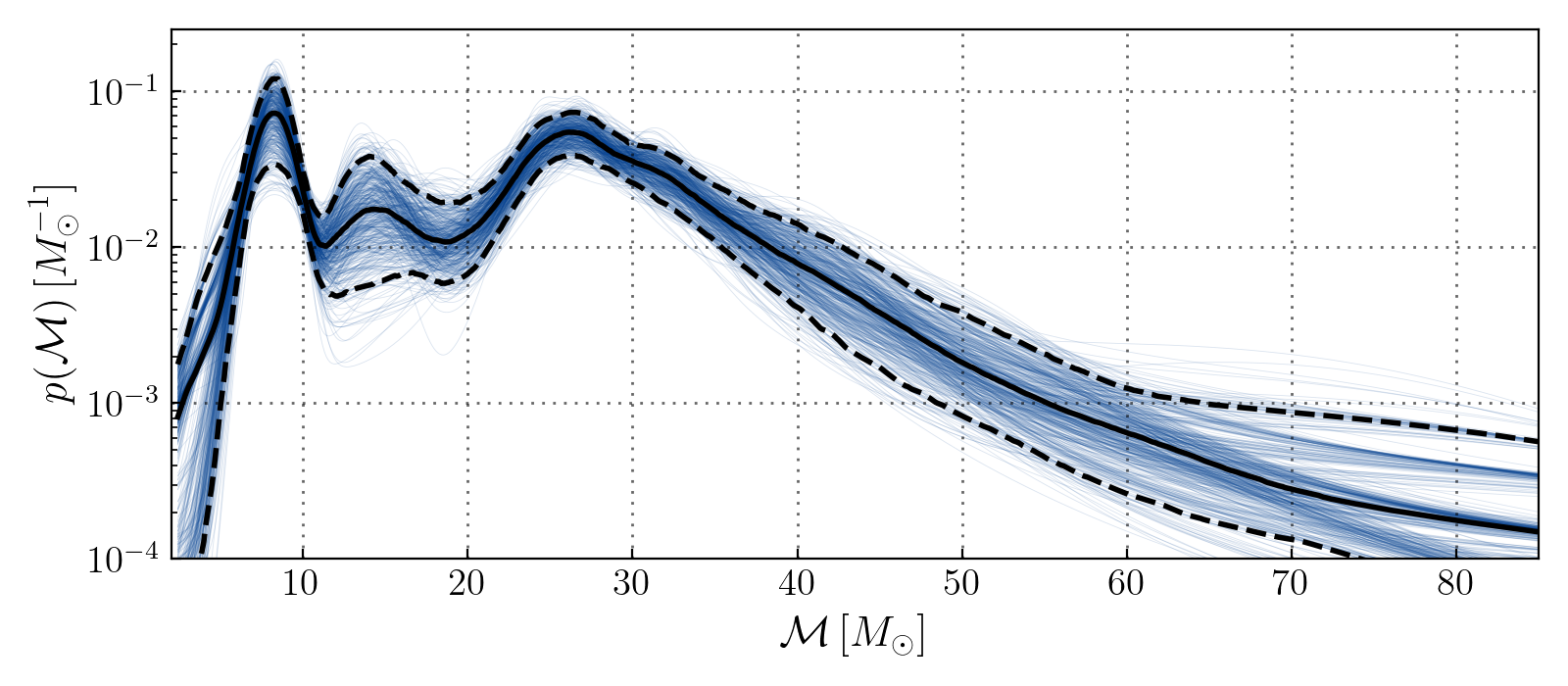}
    \caption{Application of awKDE to GWTC-3 data \cite{LIGOScientific:2021djp}: we reconstruct the source primary mass $m_1$ and chirp mass $\mathcal{M}$ distributions, including measurement uncertainties via PE samples, seeing additional features at lower masses consistent with the results in \cite{LIGOScientific:2021psn}.  (The apparently multi-modal estimate at high $\mathcal{M}$ is a by-product of the bootstrap method in regions with very low event count.)}
    \label{fig:O3bresults}
\end{figure}
As with the GWTC-2 data, we do not recover significant structure beyond the two visible peaks. The peak at $35-40\,\msun$ in primary mass ($25-30\,\msun$ in chirp mass) shows signs of asymmetry (skewness) relative to a locally Gaussian form, which may indicate a need for more complex models.

\begin{figure}[t]
\centering
\includegraphics[width=\linewidth]{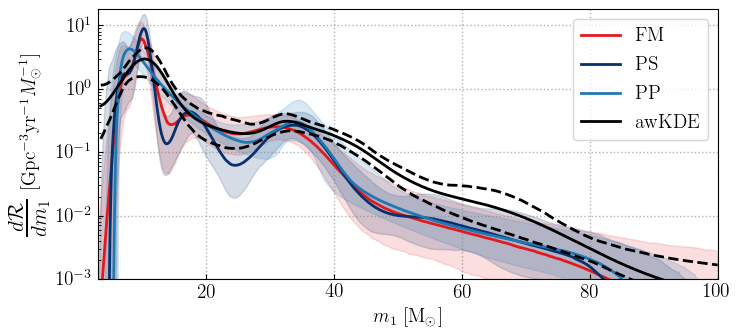}
\caption{Rate estimates using adaptive KDE and sensitive volume for BBH events in GWTC-3.  Black curves show the rate estimate using the KDE reconstruction in the top panel of Fig.~\ref{fig:O3bresults} and a fit to search sensitivity, compared with the \textsc{Flexible mixtures} (FM), \textsc{Power Law + Spline} (PS), and \textsc{Power Law + Peak} (PP) mass models, as reported in \cite{LIGOScientific:2021psn}.
}
\label{fig:O3brateestimate}
\end{figure}

\paragraph{Merger rate estimation from awKDE}
\label{sec:o3brateest}
We also estimate BBH differential merger rate over $m_1$ using our awKDE results from GWTC-3 data and an estimate of sensitive volume-time $VT$ from GWTC-3 searches, and compare the estimate with the results of \cite{LIGOScientific:2021psn}, as shown in Figure~\ref{fig:O3brateestimate}. Here we see similar features to those reported by the \textsc{Flexible mixtures}, \textsc{Power Law + Peak}, and \textsc{Power Law + Spline} models in \cite{LIGOScientific:2021psn}. 
The structure of the rate distribution obtained from KDE does differ in detail from these hierarchical analyses, probably due to our different treatment of source mass measurement uncertainties, and their use of a specific model assumption for the binary mass ratio $q$ (i.e. a power law distribution).  We also estimated $dR/dm_1$ by making a two-dimensional KDE reconstruction of detected events over $m_1, m_2$, converting to a rate distribution by dividing by $VT(m_1, m_2)$ (obviating any assumption on the distribution of $q$), and marginalizing over $m_2$: the resulting rate is consistent with our estimate in Fig.~\ref{fig:O3brateestimate}.

\subsection{2D KDE for cosmological evolution of BH mass distribution}
\begin{figure}[!th]
\centering
\includegraphics[width=0.88\linewidth, trim=2 0 10 0, clip]{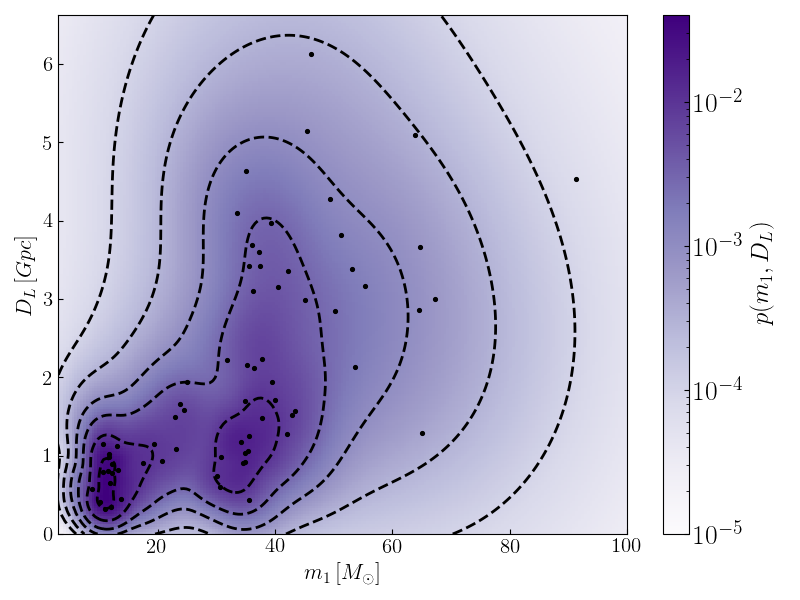}
\\
\includegraphics[width=0.91\linewidth, trim=2 0 10 0, clip]{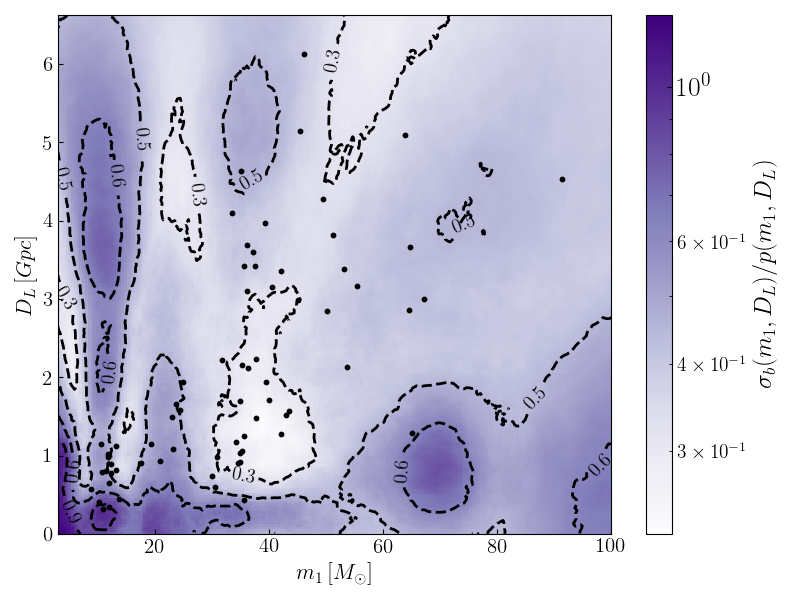}
\caption{Adaptive KDE over the source frame primary mass and luminosity distance for BBH events in GWTC-3.  The first plot shows the distribution reconstructed from 100 PE samples per event.  Features around $10\msun$ and $35\,\msun$ appear to be present consistently over different distances.  The second plot shows a relative uncertainty estimate: $\sigma_b(m_1, D_L)/p(m_1, D_L)$, where $\sigma_b = (\sigma^+ +  \sigma^-)/2$, the average of upper and lower 1-$\sigma$ bootstrap error estimates using 100 PE samples per event.  The relative error is $<1$ for most regions populated by events.}
\label{fig:twoDkdeO3b}
\end{figure}

We apply our adaptive KDE in a two-dimensional space using median PE values for $m_1$ in the source frame and luminosity distance $D_L$ for BBH events in GWTC-3~\cite{LIGOScientific:2021djp}, 
following the same procedure of leave-one-out cross-validation to determine the optimal global bandwidth and sensitivity parameter. As shown in Figure~\ref{fig:twoDkdeO3b}, the resulting two-dimensional KDE shows peaks around $10\msun$ and $35\,\msun$, similar to the one-dimensional mass KDEs and consistent with results in \cite{LIGOScientific:2021psn}; these overdensities appear to be present consistently over different distances.  The absence of detected events at low mass and higher $D_L$ is due to the selection function, i.e.\ the dependence of search sensitivity on mass.

We also show the relative uncertainty in the estimated density using bootstrap method, see bottom panel of  Figure~\ref{fig:twoDkdeO3b}, showing that over almost all the region populated by detections the uncertainty is significantly less than the density; hence the apparent peaks are unlikely to be entirely due to random counting fluctuations.

\section{Detection of BH mass function peaks with awKDE}
\label{sec:peak}
In this section we describe our algorithm \cite{peakdetectioncodelink} to detect prominent peaks in the BH mass function with awKDE. Peaks in the mass spectrum can provide hints of the formation channels responsible for the observed BBH (e.g.~\cite{Baxter:2021swn})
As shown in Figure~\ref{fig:pop_paper_models}, the two parameterised models of \cite{Abbott:2020gyp} that allow for a Gaussian peak component find such a feature around $35\,\msun$, as also visible in the awKDE reconstruction.  However, the statistical preference for such models over power-law based models with no peak component or other feature is only moderate (log10 Bayes factor $<2$ for GWTC-2 events). Here we pursue an alternative strategy by developing a general technique to identify peaks in the detected mass function, and a procedure to estimate their statistical significance.

\subsection{Peak Detection Algorithm}
We first propose an algorithm for detection of the most prominent peak in a (differentiable) one-dimensional distribution, which we then apply to our adaptive width KDEs. 
We consider the following algorithm:  
 \begin{itemize}
   \item As a pre-processing step, we cancel the maximum likelihood power-law dependence from the density estimate, in order to obtain a distribution that is as close to uniform as possible in the absence of peaks: see \ref{sec:powerlaw_opt} below.
   \item For a given distribution $\hat{f}(x)$, find all local maxima, denoted $\{x^{\mathrm{p}}_j\}$.\footnote{In practice the algorithm requires the number of local maxima, determined numerically, to not be very large, which imposes some constraint on the smoothness of the distribution.}
   \item Using a window of given size $\delta$ (the choice of $\delta$ is discussed below), evaluate the PDF on either side of the peak, $\hat{f}^{\pm}_j = \hat{f}(x^\mathrm{p}_j \pm \delta)$.  In the event that $x^{\mathrm{p}}_k \pm \delta$ is outside the range where $\hat{f}(x)$ is defined, 
   to avoid extrapolation we substitute $\hat{f}^{+}_j$ ($\hat{f}^{-}_j$) with the value of $\hat{f}$ at the highest (lowest) $x_j$ in the available range. 
   \item Compute peak heights using
   \begin{equation}
    H^{\mathrm{p}}_j = \hat{f}(x^\mathrm{p}_j) - (\hat{f}^{-}_j + \hat{f}^{+}_j) / 2.
   \end{equation}
   \item Evaluate the estimated uncertainty at each peak, $\epsilon^{\mathrm{p}}_j$, in our case defined in Eq.~\eqref{eq:error_awkde}.\footnote{If an uncertainty estimate is not available, only $H^{\mathrm{p}}_j$ can be used to define the detection statistic in the final step.}
   \item Determine the most significant peak by maximizing a detection statistic which combines $H^{\mathrm{p}}_j$ and $\epsilon^{\mathrm{p}}_j$ (we discuss the choice of detection statistic below). 
 \end{itemize}
This maximized detection statistic value will be used to distinguish distributions with significant peak features from those without such features.  
\begin{figure}[tp]
  \centering
  \includegraphics[width=0.98\linewidth]{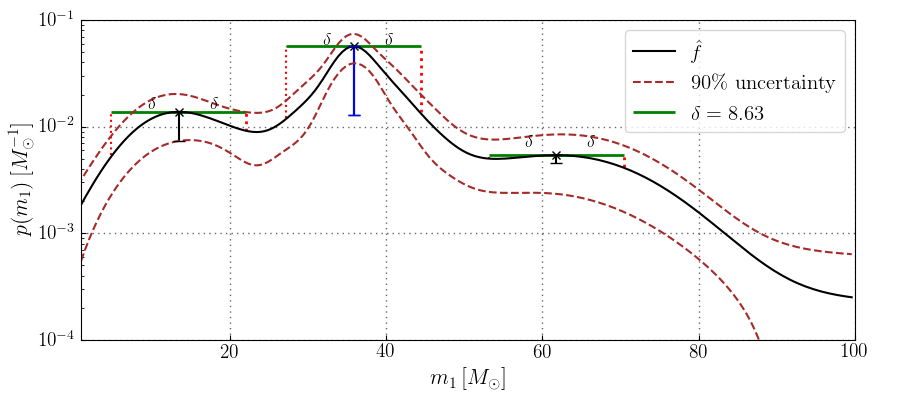}
  \caption{Peak detection algorithm for KDEs applied to BBH primary masses in GWTC-2 with an arbitrary choice of global bandwidth $h$. The solid black curve is the awKDE estimate $\hat{f}$ with the dashed brown curves indicating 5th and 95th percentile uncertainties estimated via Eq.~\eqref{eq:error_awkde}.
  The green lines indicate the window size $\delta$, 
  vertical red dotted lines indicate the construction used to measure peak height, while blue (black) vertical lines indicate the most significant (less significant) peak heights.
  (For simplicity, here we do not illustrate the pre-processing step to cancel the maximum likelihood power law dependence.)}
  \label{fig:peaks_algorithm}
\end{figure}
Figure \ref{fig:peaks_algorithm} illustrates the four steps of the algorithm subsequent to the post-processing step, applied to a set of BBH primary masses.

\subsection{Mock data sets and algorithm evaluation}
In order to evaluate and optimize the detection performance of our algorithm we created several large-scale mock datasets, each having 10,000 samples with 60 data points in each sample. Using the detection statistic values produced by the algorithm for background and signal samples, we calculate receiver operating characteristic curve (ROCs) to measure the efficiency of peak detection. 
Here, the false alarm or false positive probabilities are defined under the assumption of a truncated power-law population of mergers (accounting for the detection selection function), which we consider as representing a featureless distribution with no peak(s). 

To represent the signal hypothesis we construct a mock dataset referred to as the Peak dataset, with samples containing a mixture of a uniform (random) component and a Gaussian component.  The fraction of points drawn from the Gaussian distribution is chosen randomly between 5\%--95\%, 
the uniform component is taken to lie between $3\,\msun$--$100\,\msun$, while the mean of the Gaussian component is uniformly distributed between $8\,\msun$--$51.5\,\msun$ and its standard deviation uniformly distributed on $5\,\msun$--$10\,\msun$. 

For the background, i.e.\ no-peak hypothesis, we construct three datasets, also with 60 data points per sample, that more or less closely represent possible detected events if the true astrophysical distribution is a featureless power law. 
The first Uniform dataset contains only uniform random samples, the simplest realisation of a null case. 
The two other background datasets are drawn from a power law distribution of BBH primary and secondary masses, the \textsc{Truncated} mass model in \cite{Abbott:2020gyp} (``Model B'' in \cite{LIGOScientific:2018jsj}, see references therein). 
The model hyperparameters were estimated via Monte Carlo sampling: in our two datasets we either consider fixed hyperparameter values given by the sample medians, 
or variable hyperparameter values using available publicly released samples. 
With fixed parameter values, the primary mass power-law spectral index $\alpha$ is set to $2.21$, the lower cut-off  $m_{\rm min}$ to $5.97$, the upper cut-off $m_{\rm max}$ to $78.47$, and the mass ratio $q\equiv m_2/m_1$ follows a power-law with spectral index $\beta$ set to $1.26$. 

After generating a large number of $m_1$, $m_2$ samples from the \textsc{Truncated} model we account for the selection function. 
The probability of detection over O1, O2 and O3a data is proportional to the search sensitive volume$\times$time ($VT$), estimated as a function of $(m_1, m_2)$ via a semi-analytic calculation calibrated to search pipeline injection results
\cite{sensitivityDan} as in Section~\ref{sec:rateest}. 
Given the maximum value of $VT$ over masses, we use rejection sampling to implement the selection function. 
For the fixed hyperparameters case we obtain 60,000 data points after rejection sampling, yielding 10,000 independent samples with 60 points each. 
For the varying hyperparameters case we obtain 60 points for each hyperparameter sample; 
7460 such samples are available from the analysis of \citet{Abbott:2020gyp}. 
For GWTC-3, hyperparameter samples for the \textsc{truncated} model are not publicly available, however we were able to obtain so far unpublished results from which we use 10,000 samples,\footnote{A.~Farah, private communication.} following the same procedure as for GWTC-2.
\begin{figure}[tbp]
    \centering
    \vspace*{-0.5cm}
    \includegraphics[width=0.98\linewidth]{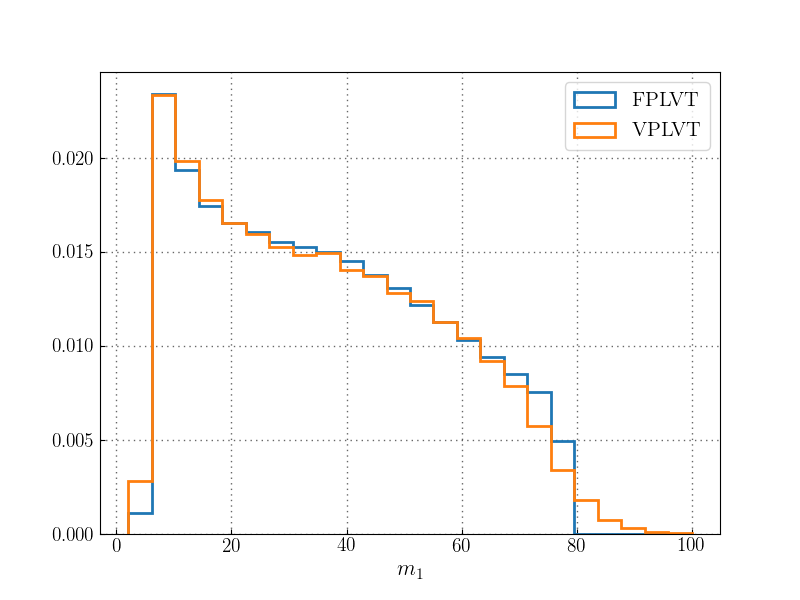}
    \caption{Distribution of background data with fixed/variable hyperparameter power law VT corrected (\mbox{FPLVT}/ \mbox{VPLVT}) datasets, based on GWTC-2 population analysis with a truncated power law model.
    }
    \label{fig:VTbackgrounddata}
\end{figure}

The resulting fixed/variable hyperparameter power-law VT corrected datasets are denoted \mbox{FPLVT}/  \mbox{VPLVT}. 
The distribution of primary masses in these power-law based datasets is shown in Fig.~\ref{fig:VTbackgrounddata} for the GWTC-2 case. 
This distribution is clearly far from uniform, however it does not have a well-defined local maximum feature apart from the cutoff at low mass. 

\begin{figure}[thp]
\centering
\includegraphics[width=0.98\linewidth]{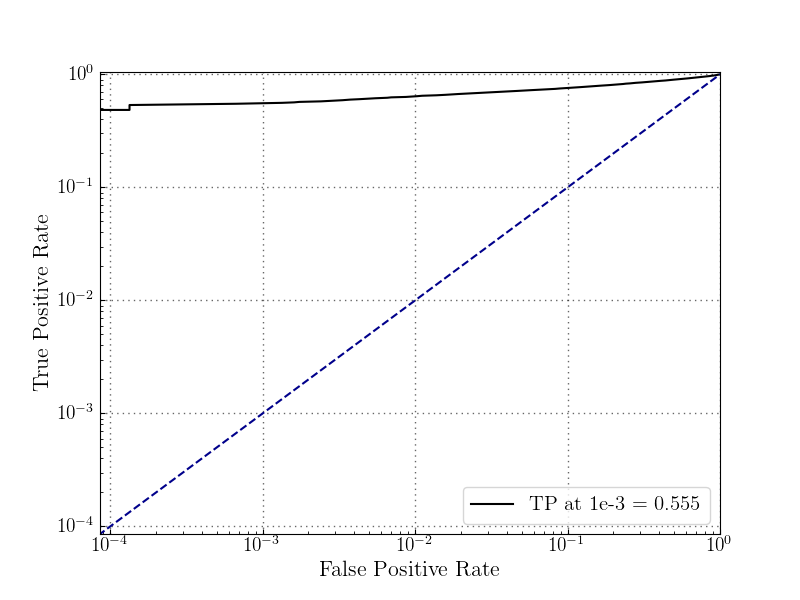}
\caption{ROC curve using the VPLVT dataset for GWTC-2 based power law samples as background and the Peak dataset as signal, for an arbitrary choice of detection statistic.  In this case the detection algorithm gives 55.5\% true positive rate at a false alarm probability of $10^{-3}$.
}
\label{fig:ROC_curve}
\end{figure}

For any specific choice of algorithm, we evaluate detection statistic values for the Peak dataset and for each of the three background datasets and then 
compute ROCs as shown in Figure~\ref{fig:ROC_curve}.  Our figure of merit is the true positive rate at a false positive rate or false alarm probability (FAP) of $10^{-3}$.  We use this criterion to tune the fixed window size $\delta$ the choice of global KDE bandwidth and the peak detection statistic. 

\subsection{Optimisation of peak detection algorithm}
\label{sec:powerlaw_opt}
As we see in Figure~\ref{fig:VTbackgrounddata}, there is a power law-like dependence in the mock datasets which can cause the peak detection algorithm to identify spurious peaks at small $m_1$ values. 
As the algorithm is designed to identify peaks above a background of nearly uniform density, in order to improve the its sensitivity we apply a pre-processing step that cancels the maximum likelihood power-law dependence from the density estimate. 
With the power-law probability density given by
\begin{equation}
f(m) = \frac{m^\gamma}{ \int_{m_{\mathrm{min}}}^{m_{\mathrm{max}}} dm\, m^\gamma},
\end{equation}
the log likelihood for $N$ events with mass values $m_i$ is
\begin{equation}
\ln \Lambda = \sum_{i=1} ^N \ln  \frac{m_{i}^{\gamma}}{ \int_{m_{\mathrm{min}}}^{m_{\mathrm{max}}} dm\, m^\gamma}. 
\end{equation}
We obtain the maximum likelihood value $\gamma_\mathrm{ML}$ value setting $\mathrm{d}\ln\Lambda/\mathrm{d}\gamma = 0$ and numerically solving the resulting equation.  Then, our pre-processing step multiplies the KDE by $m^{-\gamma_\mathrm{ML}}$ in order to suppress spurious peaks arising from a (non-uniform) power-law underlying distribution.

\subsection{Tuning the peak window $\delta$}
In our algorithm we need to choose a window size $\delta$ to compute peak heights. 
We considered both fixed (predetermined) $\delta$ choices, and values determined by the data or by the KDE construction. 

The highest figures of merit were obtained by setting $\delta = k h$, where $h$ is the standard deviation of the Gaussian kernel used to construct the initial pilot (fixed bandwidth) KDE, i.e.\ the global bandwidth, and 
$k$ is a constant factor which we took to range from 1 to 8.  Depending on other choices, we will eventually identify the value of $k$ with the highest FOM. 
Given our method of selecting $h$, the choice $\delta = k h$ ensures that the window size is automatically scaled to the variance of the data, 
whereas any fixed $\delta$ value would have to be retuned when applying the method to other physical situations.

\subsection{Tuning the global KDE bandwidth} 
In order to choose the global bandwidth for the pilot KDE we do not consider leave-one-out cross validation, as for the 
10,000 samples of our background data this would be too computationally expensive.  Also, the optimal bandwidth for accurate reconstruction of the mass distribution is not necessarily most suited to detection of peaks.  Instead, we construct an awKDE for each sample, fixing $\alpha = 1$, for several different choices of bandwidth between $0.1$ and $0.5$ times the standard deviation of the data points, uniformly spaced in $\log(h)$. We restrict the maximum bandwidth 
to $0.5$ in units of the data standard deviation  since if the bandwidth is larger 
the resulting KDE is necessarily very smooth, without smaller scale features. This affects the goal of our algorithm to identify datasets which have a peak feature smaller than the total overall extent of the distribution, versus those without such a feature. Out of these different bandwidths for a given sample, we select the one for which the maximum peak detection statistic is highest. 

\subsection{Range of KDE evaluation and boundary conditions} 
In order to check any specific trends in our algorithm we plotted several diagnostic tests including peak height, peak detection statistic and peak location for our mock datasets. We found that some spurious peaks were generated in peak datasets due to the falloff of the KDE towards the edges of the range, a known artifact resulting from the absence of points outside the range \cite{lscsoftboundedKDE,Hoy:2020vys}.  To suppress these spurious peaks we use a bounded KDE, effectively reflecting the data points about the edges of the range, and we check that this choice increases the detection figure of merit.  
The choice of range of values over which to evaluate the KDE also affects the detection algorithm: we find the best performance for a KDE range which extends the range of the data points by 10\% below (above) the minimum (maximum) points in a given sample. 

\subsection{Tuning choice of peak detection statistic}
We lastly need to specify the detection statistic which determines the most significant peak in each awKDE.  We considered various possibilities: the simplest choice would be to use the peak height $H^{\mathrm{p}}_j$, however we also investigated combinations of the peak height and the error estimate $\epsilon_\mathrm{KDE}(x^\mathrm{p}_j)$ at the peak as in Eq.~\eqref{eq:error_awkde}. Considering the ratio $H^{\mathrm{p}}_j/ \epsilon_\mathrm{KDE}(x^\mathrm{p}_j)$ as detection statistic, it has higher FOM than simply the peak height.  However, the error estimate $\epsilon_\mathrm{KDE}(x^\mathrm{p}_j)$ reduces sharply with increasing bandwidth, leading to some spurious peaks in our background datasets with small height but smaller estimated errors. 

Finally, to avoid such artefacts at high bandwidth we consider the modified uncertainty estimate of Eq.~\eqref{eq:sigmahat}, which better agrees with the uncertainty obtained from bootstrap resampling.  Using the statistic $H^{\mathrm{p}}_j/\hat{\epsilon}_\mathrm{KDE}(x^\mathrm{p}_j)$ the FOM increased further, thus this expression is our final choice. 

\subsection{Optimized detection choice}
After the tuning steps described above, our choice of peak detection statistic is the ratio $H^{\mathrm{p}}_j/\hat{\epsilon}_\mathrm{KDE}(x^\mathrm{p}_j)$: then for the FPLVT background dataset $\delta = 3 h$ gives the highest FOM, whereas for the VPLVT dataset $\delta = 5 h$ gives the highest FOM (nearby choices give very similar performance). Hence we use $\delta = 4 h$
to compute the detection statistic for the real dataset consisting of observed GW signals from BBH merger.

\subsection{Peak detection for GWTC-2 BBH events}
We apply our optimized detection algorithm to the 44 high-significance BBH detections in GWTC-2 as used in \cite{Abbott:2020gyp}, finding a peak detection statistic of $2.85$. To assess the significance of this value we construct comparable background samples, each having 44 data points, analogously to the FPLVT (VPLVT) mock data described in the previous section; as before we obtain 10000 (7640) samples and apply the same algorithm to compute their detection statistics. 
  
We then compute the FAP for the detected events up to O3a using these background datasets, obtaining a value 0.0001 ($3.7\sigma$ significance) for the FPLVT background and a value of 0.00013 ($3.6\sigma$) for the VPLVT background.\footnote{These results are obtained using the pre-processing procedure of Sec.\ref{sec:powerlaw_opt}; in our previous analysis without this step we obtained FAP values 0.0005 ($3.3\sigma$) for the FPLVT background and 0.0016 ($2.9\sigma$) for VPLVT background.} 

So far, we have not accounted for mass measurement error in constructing background samples.  We expect that such errors would not cause more significant peaks to appear in the KDE of BBH primary masses, however we investigate this by simulating errors of similar magnitude to those in the primary masses of GWTC-2 events \cite{LIGOScientific:2020ibl}. 
We approximate the distribution of errors in $m_1$ as log-normal, and find a representative standard deviation of $0.24$ by considering the published 90\% credible intervals. 
Applying errors with this distribution to the VPLVT background dataset, we recompute the detection statistic values and obtain a FAP for the detected events up to O3a 
of 0.0017 ($2.9\sigma$).  Hence such errors do not strongly impact our conclusions. 

Thus, the apparent feature in GWTC-2 BBH primary masses is very unlikely to have originated from random fluctuations, if the true underlying distribution was a power law.  This result strengthens the motivation to search for possible astrophysical explanations of such features. 

\subsection{Peak detection for GWTC-3 BBH events}
We further applied the detection algorithm to 69 high-significance (false alarm rate below 0.25/yr) BBH events in GWTC-3~\cite{LIGOScientific:2021djp}, finding a peak detection statistic of $3.36$.  Using FPLVT (VPLVT) mock data with 10000 samples, each having 69 data points, we compute their detection statistics.  We then compute the FAP for the detected events in GWTC-3 using these background datasets, obtaining values 0.0001 ($3.7\sigma$) for the FPLVT background and 0.0012 ($3.0\sigma$) for the VPLVT background.  Following the same procedure as for GWTC-2 to approximate the distribution of errors in $m_1$ as log-normal, we find a representative standard deviation of $0.23$ by considering the published 90\% credible intervals. 
Applying errors with this distribution to the VPLVT background dataset, we recompute the detection statistic values and obtain a FAP for GWTC-3 events of 0.0103 ($2.3\sigma$).  

This apparently reduced significance for GWTC-3 events against VPLVT backgrounds, despite the higher peak detection statistic, reflects a difference in the \textsc{truncated} hyperparameter samples used for the background: specifically a more negative power-law index for the primary mass in GWTC-3 as compared to GWTC-2, leading to spurious peaks with somewhat higher detection statistics despite the steps taken in our algorithm. 
Thus, it is likely that the algorithm could be further refined to better reject steep power law background distributions. Our astrophysical conclusion that the primary BH mass distribution contains a peak-like feature strongly inconsistent with a power law is not affected.

\section{Conclusions}

GW observations from compact binary mergers detected in LVK data provide key astrophysical constraints, by allowing the comparison of theoretical models to the observed population of BBH.  Many techniques to measure this population rely on a predetermined functional model form ({e.g.}, \cite{Fishbach:2017zga}, \cite{Talbot:2018cva}), which cannot react to surprises without human intervention.  In this work we provide a new, computationally cheap method for measuring the population with weak assumptions, allowing for the automated discovery of new population features; it will thus help to answer questions regarding mass gaps, sub-populations, and other signatures of BH formation channels as the number of detections increases.

Our fast and transparent method uses adaptive bandwidth KDE (awKDE) \cite{awkdecodelink} to estimate or reconstruct the population distribution of compact binary mergers detectable by the global gravitational-wave interferometer network. The resulting estimates can serve as sanity checks on  computationally more expensive Bayesian hierarchical inferences on binary merger populations, and may also be able to bring to light features that are not present in commonly used parameterized models. The method can straightforwardly incorporate uncertainties in individual source parameters and yield estimates of astrophysical merger rates.  We specifically demonstrated its use for binary black hole (BBH) merger distributions over primary mass, chirp mass and cosmological distance.

In addition, we introduce a robust peak detection algorithm \cite{peakdetectioncodelink} for smooth one-dimensional distributions, such as those produced by awKDE, 
and apply the algorithm to BBH primary masses from the second LIGO-Virgo catalog of transient GW sources, obtaining significant evidence for the presence of a peak feature in the mass distribution as opposed to a simple power law.  Possible future developments of this algorithm include extensions to detect gaps or dips in the distribution, to detect multiple features, or to better account for dependence of the detection probability on binary masses (or other parameters). 

\section*{Acknowledgements}
The authors thank Amanda Farah for sharing unpublished population analysis results for the \textsc{Truncated} model using GWTC-3 data. This work has received financial support from Xunta de Galicia (Centro singular de investigación de Galicia accreditation 2019-2022), by European Union ERDF, and by the ``María de Maeztu'' Units of Excellence program MDM-2016-0692 and the Spanish Research State Agency.  DW thanks the NSF (PHY-1912649) for support. This material is based upon work supported by NSF's LIGO Laboratory which is a major facility fully funded by the National Science Foundation. The authors are grateful for computational resources provided by the LIGO Laboratory and supported by National Science Foundation Grants PHY-0757058 and PHY-0823459. This research has made use of data, software and/or web tools obtained from the Gravitational Wave Open Science Center (https://www.gw-openscience.org/), a service of LIGO Laboratory, the LIGO Scientific Collaboration and the Virgo Collaboration. LIGO Laboratory and Advanced LIGO are funded by the United States National Science Foundation (NSF) as well as the Science and Technology Facilities Council (STFC) of the United Kingdom, the Max-Planck-Society (MPS), and the State of Niedersachsen/Germany for support of the construction of Advanced LIGO and construction and operation of the GEO600 detector. Additional support for Advanced LIGO was provided by the Australian Research Council. Virgo is funded, through the European Gravitational Observatory (EGO), by the French Centre National de Recherche Scientifique (CNRS), the Italian Istituto Nazionale di Fisica Nucleare (INFN) and the Dutch Nikhef, with contributions by institutions from Belgium, Germany, Greece, Hungary, Ireland, Japan, Monaco, Poland, Portugal, Spain.

\bibliographystyle{apsrev4-1}
\bibliography{reference}
\end{document}